\documentclass[usegraphicx,psfig,letterpaper]{mn2e}

\usepackage{amssymb}
\usepackage{times}
\usepackage{psfig}
\usepackage{amsmath}
\def\ltsima{$\; \buildrel < \over \sim \;$}
\def\simlt{\lower.5ex\hbox{\ltsima}}
\def\gtsima{$\; \buildrel > \over \sim \;$}
\def\simgt{\lower.5ex\hbox{\gtsima}}

\newread\epsffilein    % file to \read
\newif\ifepsffileok    % continue looking for the bounding box?
\newif\ifepsfbbfound   % success?
\newif\ifepsfverbose   % report what you're making?
\newdimen\epsfxsize    % horizontal size after scaling
\newdimen\epsfysize    % vertical size after scaling
\newdimen\epsftsize    % horizontal size before scaling
\newdimen\epsfrsize    % vertical size before scaling
\newdimen\epsftmp      % register for arithmetic manipulation
\newdimen\pspoints     % conversion factor
\def\epsfbox#1{\global\def\epsfllx{72}\global\def\epsflly{72}%
   \global\def\epsfurx{540}\global\def\epsfury{720}%
   \def\lbracket{[}\def\testit{#1}\ifx\testit\lbracket
   \let\next=\epsfgetlitbb\else\let\next=\epsfnormal\fi\next{#1}}%
\def\epsfgetlitbb#1#2 #3 #4 #5]#6{\epsfgrab #2 #3 #4 #5 .\\%
   \epsfsetgraph{#6}}%
\def\epsfnormal#1{\epsfgetbb{#1}\epsfsetgraph{#1}}%
\def\epsfgetbb#1{%
\openin\epsffilein=#1
\ifeof\epsffilein\errmessage{I couldn't open #1, will ignore it}\else
   {\epsffileoktrue \chardef\other=12
    \def\do##1{\catcode`##1=\other}\dospecials \catcode`\ =10
    \loop
       \read\epsffilein to \epsffileline
       \ifeof\epsffilein\epsffileokfalse\else
          \expandafter\epsfaux\epsffileline:. \\%
       \fi
   \ifepsffileok\repeat
   \ifepsfbbfound\else
    \ifepsfverbose\message{No bounding box comment in #1; using defaults}\fi\fi
   }\closein\epsffilein\fi}%
\def\epsfsetgraph#1{%
   \epsfrsize=\epsfury\pspoints
   \advance\epsfrsize by-\epsflly\pspoints
   \epsftsize=\epsfurx\pspoints
   \advance\epsftsize by-\epsfllx\pspoints
   \epsfxsize\epsfsize\epsftsize\epsfrsize
   \ifnum\epsfxsize=0 \ifnum\epsfysize=0
      \epsfxsize=\epsftsize \epsfysize=\epsfrsize
     \else\epsftmp=\epsftsize \divide\epsftmp\epsfrsize
       \epsfxsize=\epsfysize \multiply\epsfxsize\epsftmp
       \multiply\epsftmp\epsfrsize \advance\epsftsize-\epsftmp
       \epsftmp=\epsfysize
       \loop \advance\epsftsize\epsftsize \divide\epsftmp 2
       \ifnum\epsftmp>0
          \ifnum\epsftsize<\epsfrsize\else
             \advance\epsftsize-\epsfrsize \advance\epsfxsize\epsftmp \fi
       \repeat
     \fi
   \else\epsftmp=\epsfrsize \divide\epsftmp\epsftsize
     \epsfysize=\epsfxsize \multiply\epsfysize\epsftmp   
     \multiply\epsftmp\epsftsize \advance\epsfrsize-\epsftmp
     \epsftmp=\epsfxsize
     \loop \advance\epsfrsize\epsfrsize \divide\epsftmp 2
     \ifnum\epsftmp>0
        \ifnum\epsfrsize<\epsftsize\else
           \advance\epsfrsize-\epsftsize \advance\epsfysize\epsftmp \fi
     \repeat     
   \fi
   \ifepsfverbose\message{#1: width=\the\epsfxsize, height=\the\epsfysize}\fi
   \epsftmp=10\epsfxsize \divide\epsftmp\pspoints
   \newcount\figskipcount
      \message{#1 \the\epsfysize  }
   \vbox to\epsfysize{\vfil\hbox to\epsfxsize{%
      \includegraphics{#1}%
      \hfil}}%
\epsfxsize=0pt\epsfysize=0pt}%

{\catcode`\%=12 \global\let\epsfpercent=%\global\def\epsfbblit{%BoundingBox}}%
\long\def\epsfaux#1#2:#3\\{\ifx#1\epsfpercent
   \def\testit{#2}\ifx\testit\epsfbblit
      \epsfgrab #3 . . . \\%
      \epsffileokfalse
      \global\epsfbbfoundtrue
   \fi\else\ifx#1\par\else\epsffileokfalse\fi\fi}%
\def\epsfgrab #1 #2 #3 #4 #5\\{%
   \global\def\epsfllx{#1}\ifx\epsfllx\empty
      \epsfgrab #2 #3 #4 #5 .\\\else
   \global\def\epsflly{#2}%
   \global\def\epsfurx{#3}\global\def\epsfury{#4}\fi}%
\def\epsfsize#1#2{\epsfxsize}
\begin{document}
\title[A Hubble Space Telescope study of SCUBA
submillimetre galaxies]
 {A Hubble Space Telescope study of SCUBA submillimetre galaxies}
\author[O. Almaini et al.]
{
O.~Almaini,$^{1}$ 
J.S.~Dunlop,$^{2}$ C.J.~Conselice,$^{3}$
T.A.~Targett,$^{2}$ R.J.~McLure$^2$
\\
$^1$ School of Physics \& Astronomy, University of Nottingham,
University Park, Nottingham NG7 2RD 
\\ $^2$ Institute for Astronomy,
University of Edinburgh, Royal Observatory, Blackford Hill, Edinburgh
EH9 3HJ \\ $^3$ California Institute of Technology, Pasadena, CA
91125, USA
}
\date{MNRAS, submitted}
\maketitle

\begin{abstract}
We present high-resolution imaging of a sample of 10 SCUBA
submillimetre galaxies, obtained using the ACS camera on the Hubble
Space Telescope. We find that the majority show compact, disturbed
morphologies. Using quantitative morphological classification, we find
that at least 6 are classified as major mergers.  Simulations suggest
that the morphological parameters are unlike local spirals and
ellipticals, but similar to those of local ULIRGs.  Compared to
Lyman-break galaxies, the submillimetre-selected galaxies are on
average more asymmetric, but also significantly more concentrated in
their light distributions.  This is consistent with a higher fraction
of the stellar mass being contained within a central spheroid
component.  Despite their different morphologies, we find that submm
galaxies are similar in size to  luminous ($\simgt L^*$)
Lyman-break galaxies, with half-light radii in the range
$2.8-4.6$~kpc.
\end{abstract}


\section{Introduction}

Surveys in the submm waveband have revealed a population of highly
luminous, dust enshrouded galaxies that appear to contribute a large
fraction of the star-formation at $z>2$ (Smail et al. 1997; Hughes et
al.  1998; Barger et al. 1998, Eales et al. 1999).  Originally
discovered using the SCUBA array at the James Clerk Maxwell Telescope,
these galaxies have been heralded by many as the discovery of the
major epoch of dust-enshrouded spheroid formation (Lilly et al. 1999,
Dunlop 2001, Granato et al. 2001).  In terms of their bolometric power
output (typically $>10^{12} L_{\odot}$), the submm sources appear to
be high-z analogues of local Ultra-Luminous Infrared Galaxies (ULIRGs;
Sanders \& Mirabel 1996).  The major difference, however, is their
space density. At high redshift the submm galaxies may dominate the
cosmic star-formation rate, while locally ULIRGs are rare and unusual
phenomena.

Beyond their initial discovery, progress in understanding these
galaxies has been difficult.  This is due to an unfortunate
combination of their optical faintness and the relatively large ($\sim
10$ arcsec) SCUBA beam, which often makes unambiguous identification
impossible.  This deadlock has been partially overcome with deep radio
observations at the Very Large Array (VLA), which have given precise
locations for a significant sample of submm galaxies for the first
time (Ivison et al. 2002, Chapman et al. 2003a). Follow-up
spectroscopy has produced the first reliable $N(z)$ estimate, showing
a median redshift of $z=2.4$ with only a small fraction (a few per
cent) of galaxies at $z<1$ (Chapman et al. 2003a).

There have been suggestions that many submm sources could be powered
by AGN (e.g. Almaini, Lawrence \& Boyle 1999), but the failure to
detect significant numbers as luminous X-ray sources suggests that the
AGN-dominated fraction is likely to be small (Fabian et al.\ 2000, 
Severgnini et al. 2000, Almaini et al. 2003). Nevertheless, a large
fraction appear to host moderate-luminosity AGN activity (Alexander et
al. 2003; Alexander et al. 2005), which may indicate that we are observing
the joint-formation epoch of black holes and spheroids.

Despite the observational progress, current semi-analytic models have
great difficulty producing so many dusty, luminous objects at these
redshifts without resorting to extreme model parameters (Baugh et
al. 2005). Morphological information offers a potential additional
test of these models.  For example, do they have structural parameters
that are similar to local galaxies, or do they look more like ULIRGs?
Are they compact or extended? What fraction is undergoing merging,
and how do they differ in their structural parameters from more
`typical' galaxies at $z>2$, as detected by photometric selection
techniques (Steidel et al. 1996)?

Ground-based morphological classification has been difficult, since
the optically-identified submm galaxies are typically very faint and
compact (Smail et al. 2004), but studies with the Hubble Space
Telescope have revealed evidence for disturbed, multi-component
morphologies (Conselice et al. 2003b, Smail et al. 2004, Pope et
al. 2005). There have also been claims that submm sources are
significantly larger than more typical field galaxies at high redshift
(Chapman et al. 2003b, Smail et al. 2004, Pope et al. 2005), which
would add weight to the hypothesis that these are massive galaxies in
formation (Swinbank et al. 2004, Tecza et al. 2004).

In this paper we perform an independent morphological analysis on a
flux-limited sample of SCUBA submillimetre galaxies from a contiguous
survey field. We attempt to overcome the bias inherent in some
previous studies by also seeking counterparts of submm sources which
do not have a clear radio identification.

Our HST imaging is complemented by deep K-band imaging from the NIRI
instrument on Gemini.  In a companion paper (Targett et al. 2005, in
preparation) we explore the K-band light profiles and colour gradients
of submm galaxies in more detail.

The layout of this paper is as follows.  In Section 2 we present the
sample and observational data. In Section 3 we discuss the optical/IR
identification of the submm sources, which we separate into 3
categories (secure, likely and unknown).  In Section 4 we present a
quantitative morphological analysis, followed by a study of galaxy
sizes in Section 5. In Section 6 we briefly discuss the properties of
individual galaxies before presenting our conclusions in Section
7. All optical/IR magnitudes are in the Vega system unless stated
otherwise.  Cosmological parameters $h=0.7,\Omega_m=0.3,
\Omega_{\Lambda}=0.7$ are assumed throughout.

\begin{figure*}
\centerline{\psfig{figure=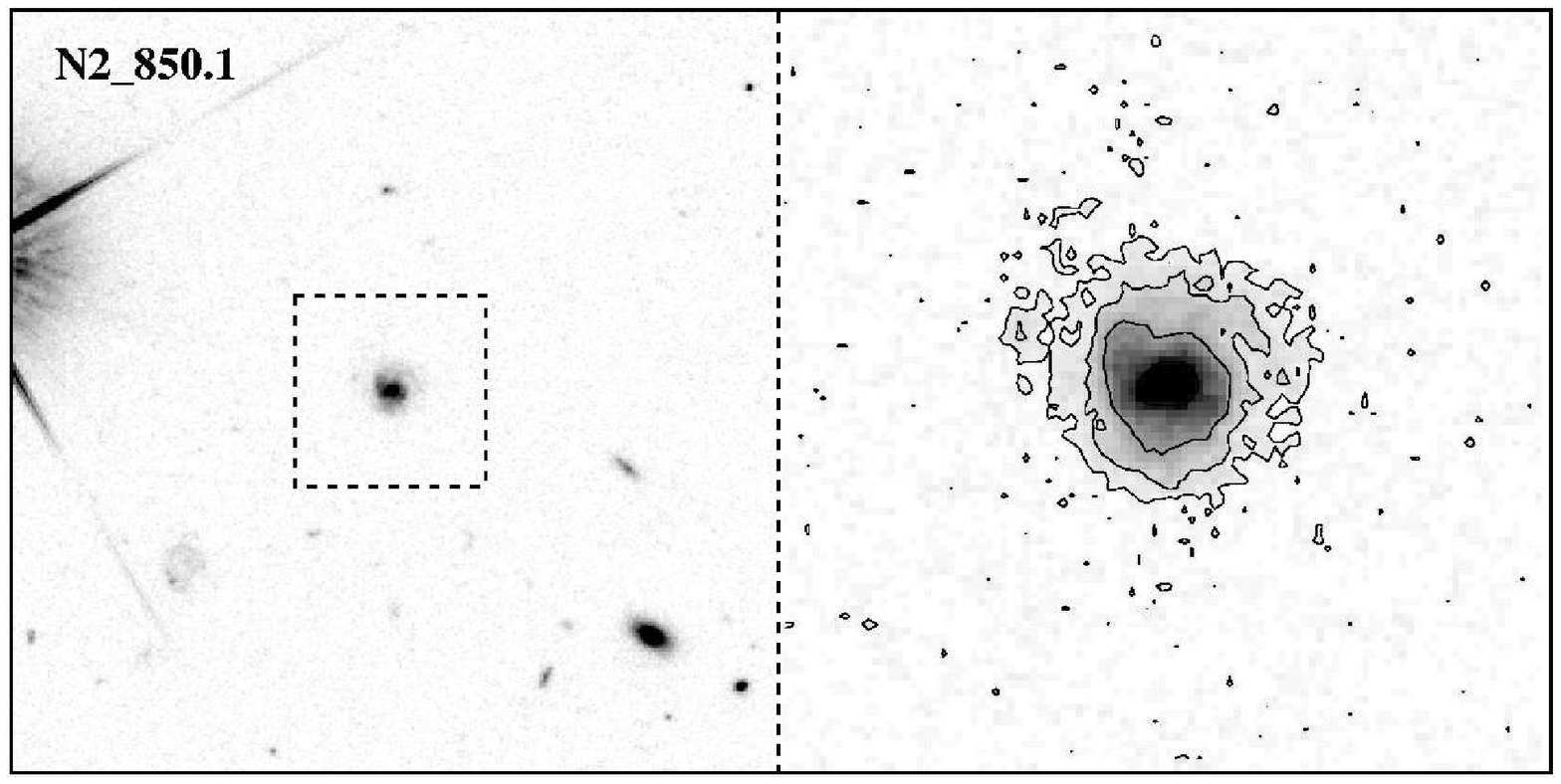,width=0.5\textwidth,angle=0}
\psfig{figure=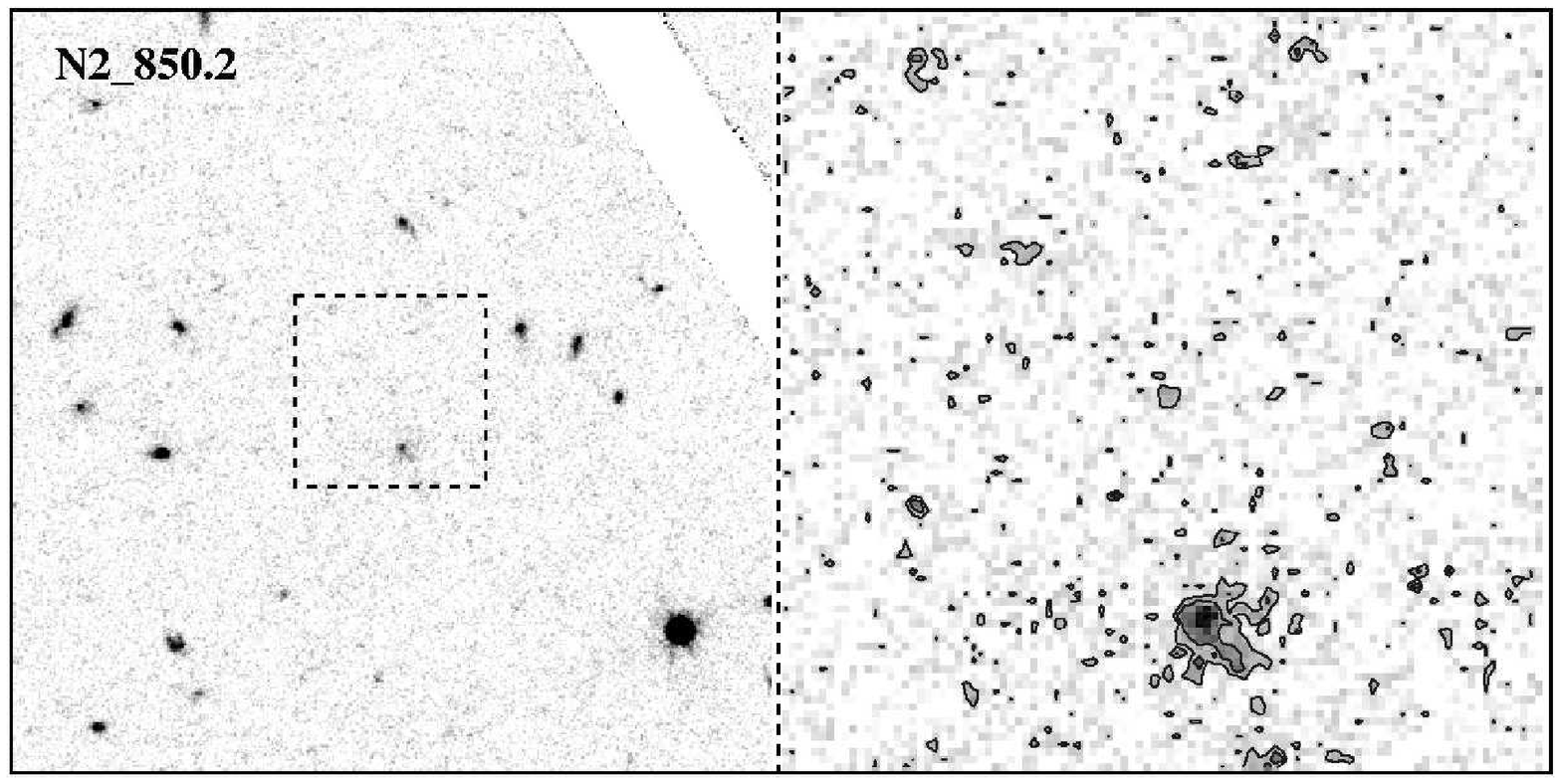,width=0.5\textwidth,angle=0}}
\centerline{\psfig{figure=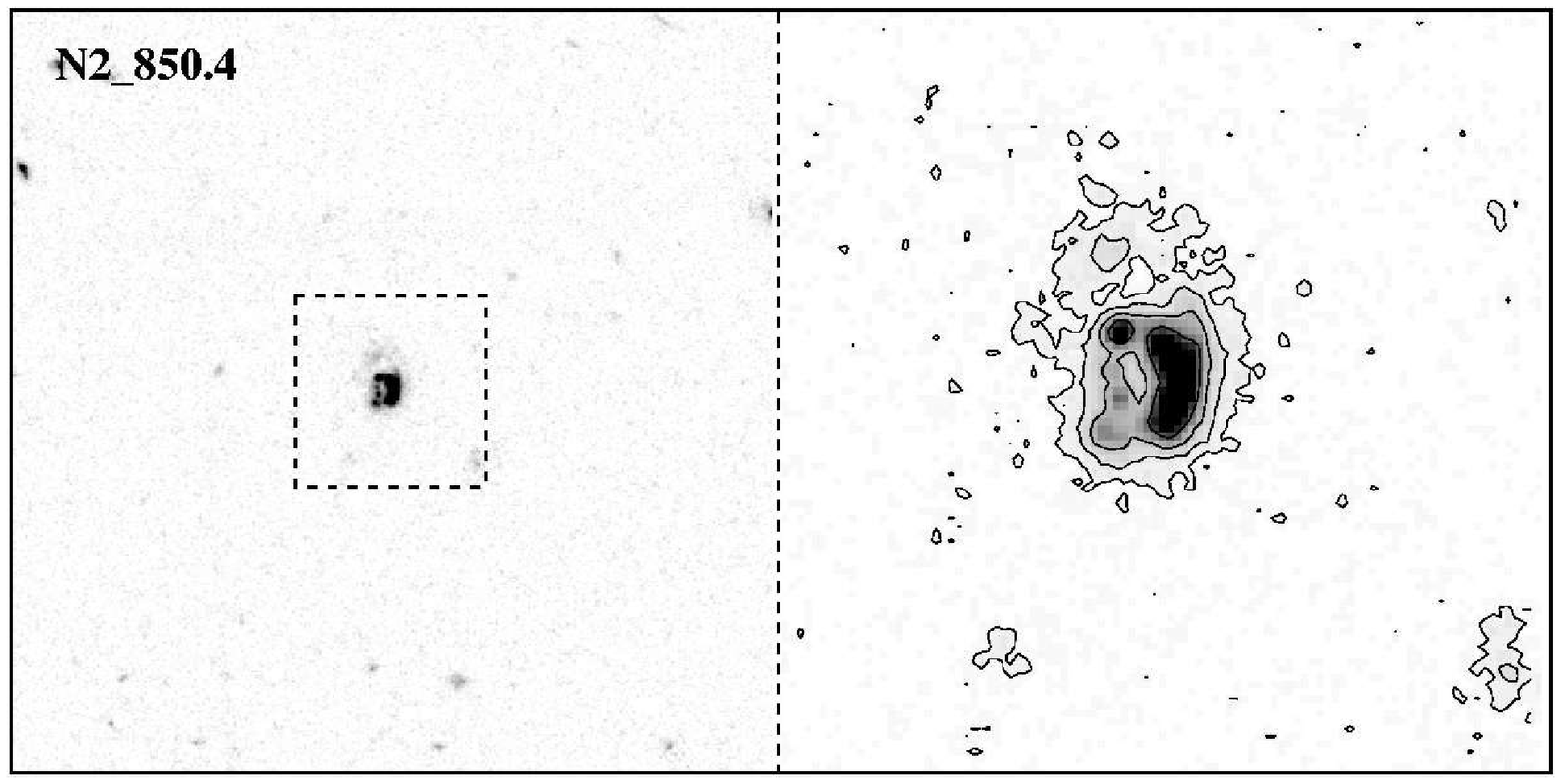,width=0.5\textwidth,angle=0}
\psfig{figure=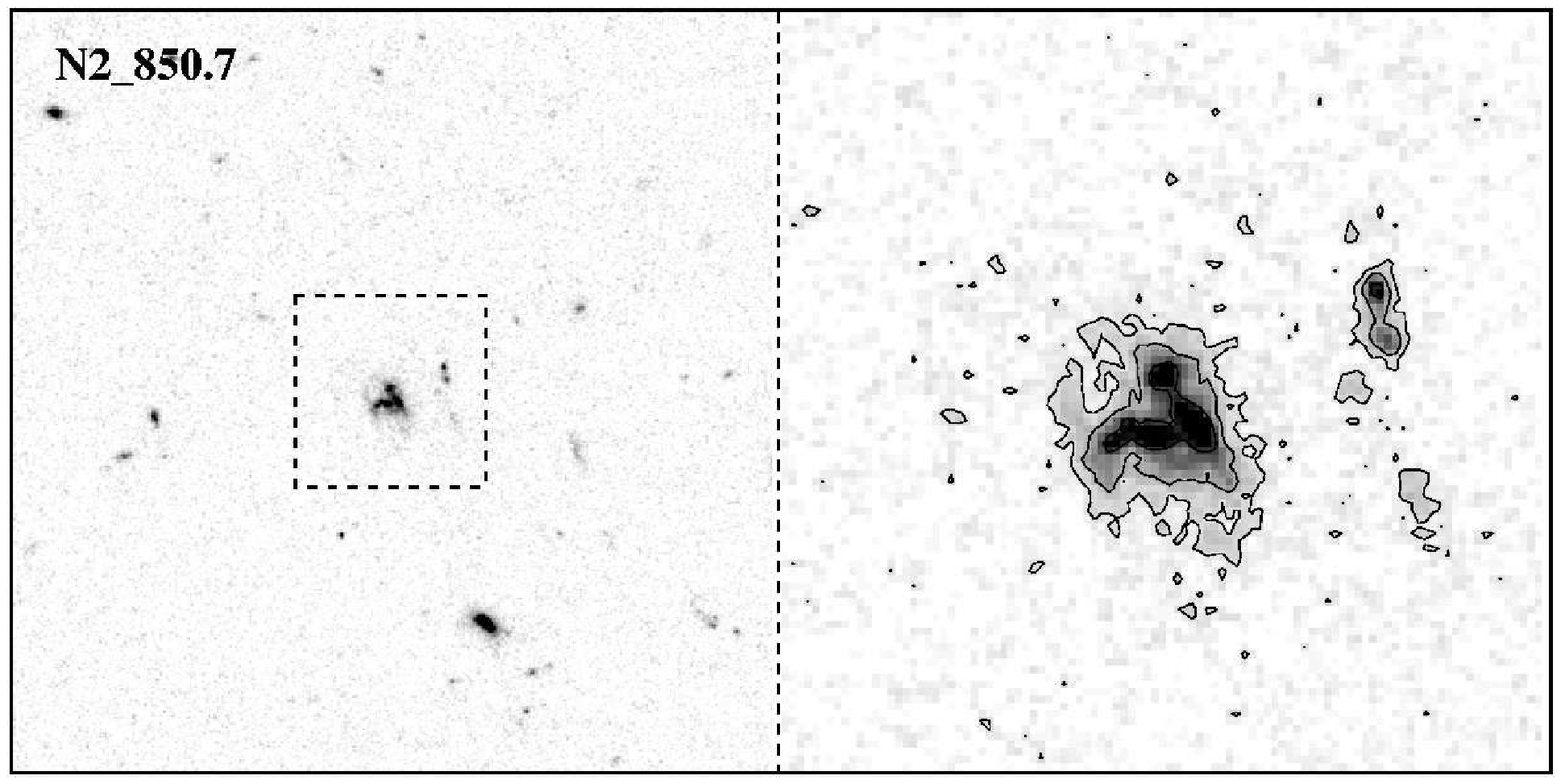,width=0.5\textwidth,angle=0}}
\centerline{\psfig{figure=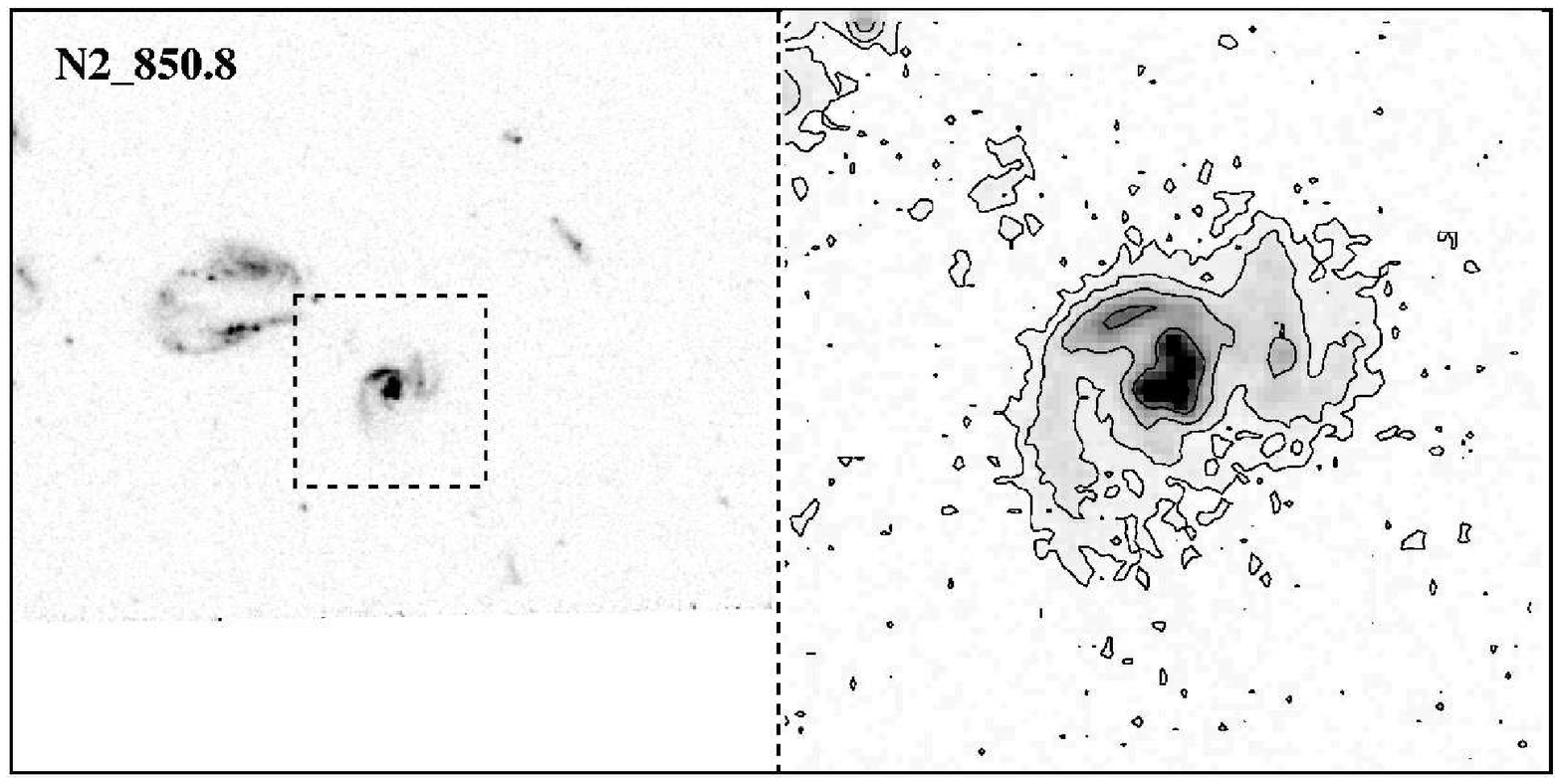,width=0.5\textwidth,angle=0}
\psfig{figure=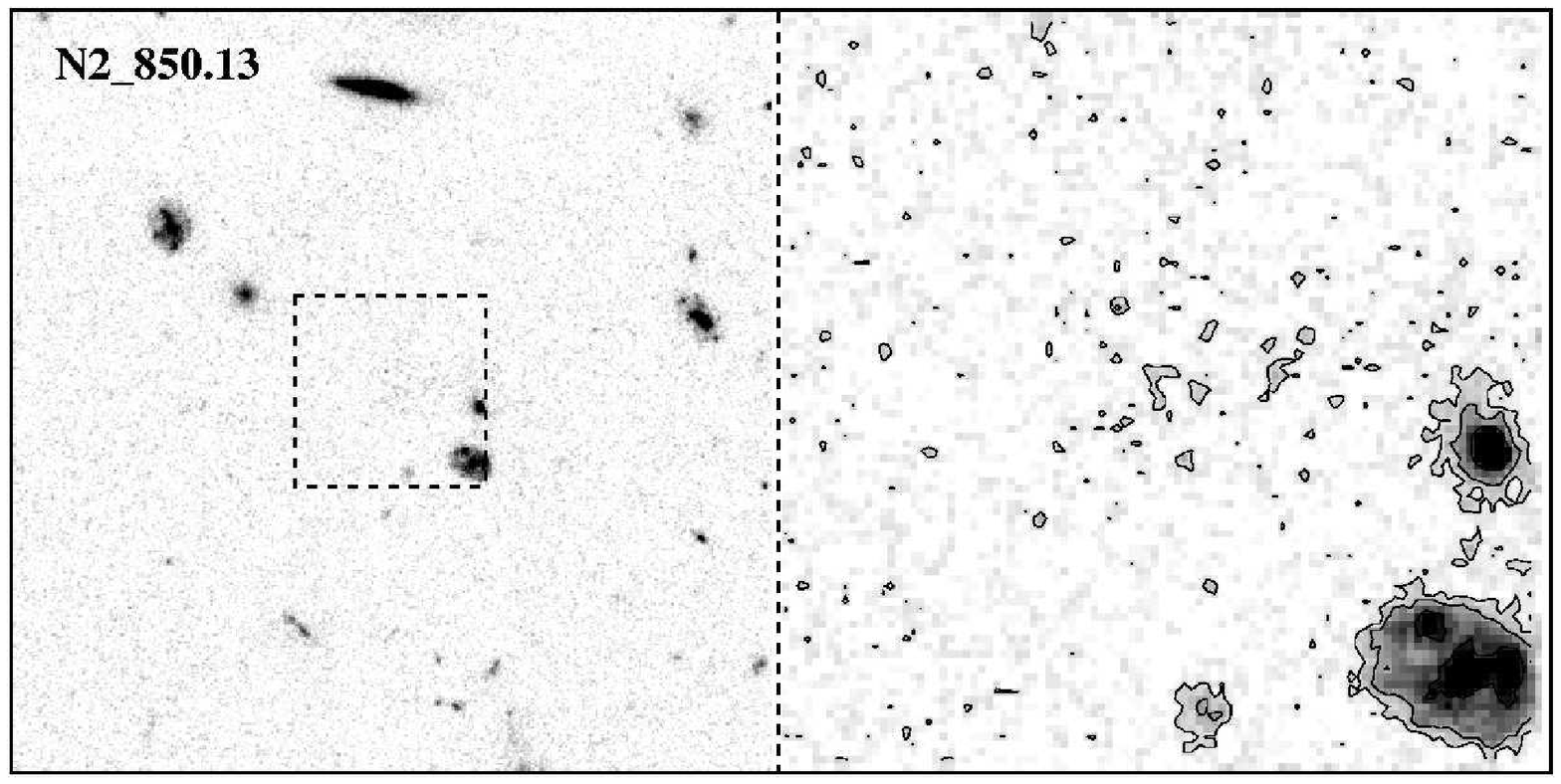,width=0.5\textwidth,angle=0}}
\caption{HST-ACS images of the 6 submm sources with {\it secure} radio
identifications, taken in the F814 filter and centred on the radio
counterpart. The $20\times 20$ arcsec postage stamps (left panels) permit a
comparison with the work of Ivison et al. (2002). The  $5\times 5$
arcsec blow-up regions are shown with contours to reveal low surface-brightness
features. The contour are separated by intervals of 1
magnitude/arcsec$^2$.}
\end{figure*}

\begin{figure*}
\centerline{\psfig{figure=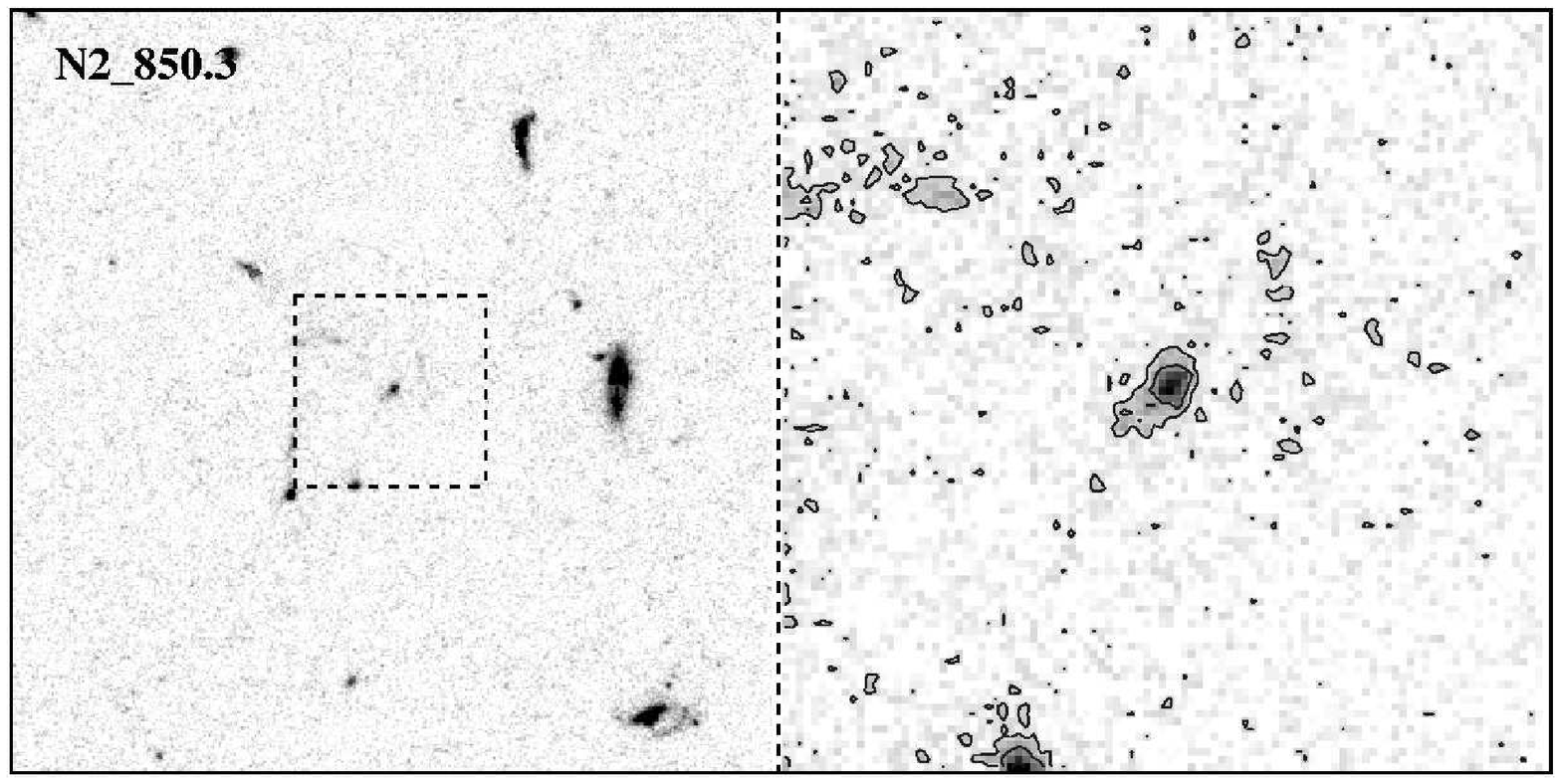,width=0.5\textwidth,angle=0}
\psfig{figure=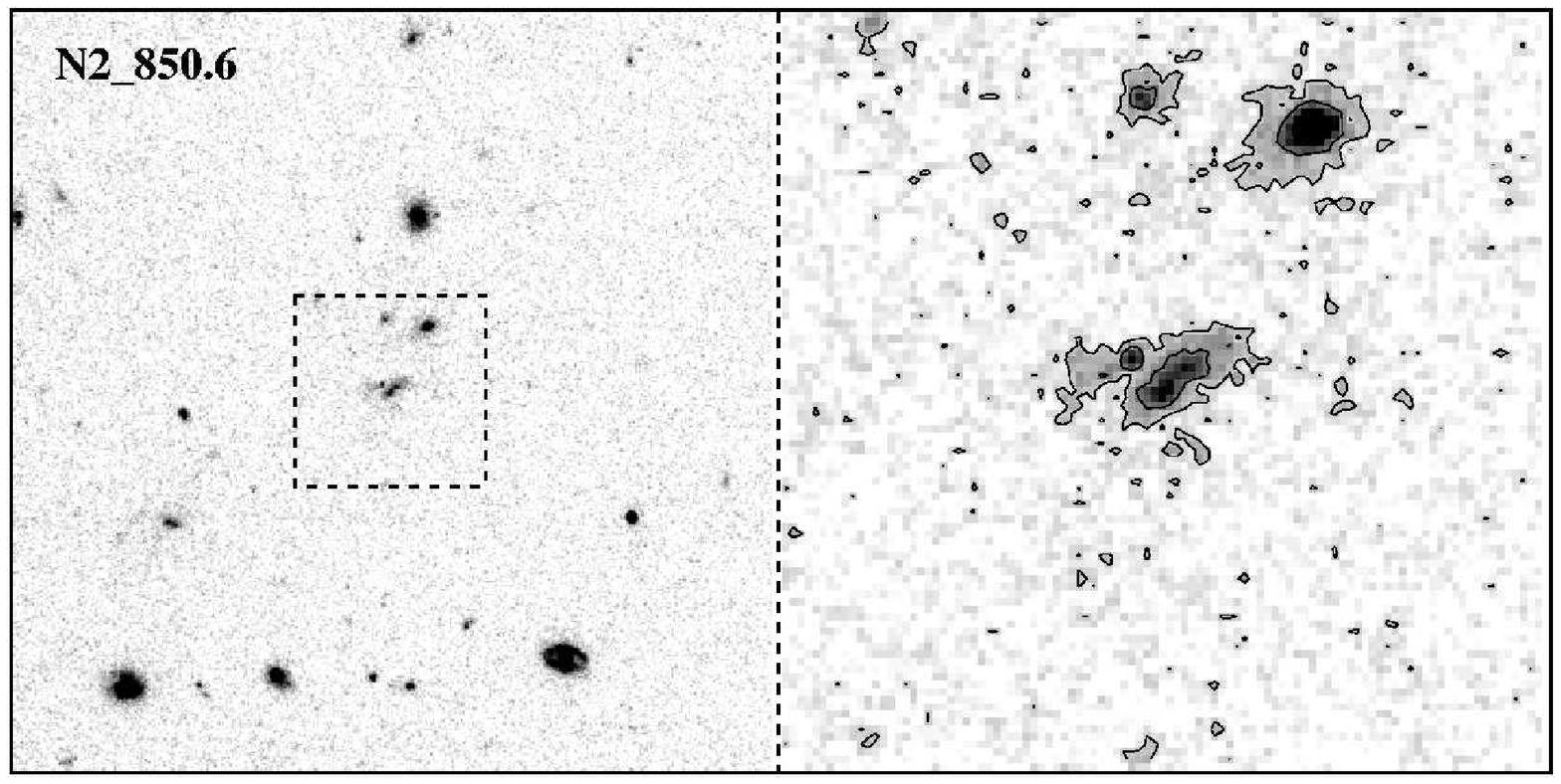,width=0.5\textwidth,angle=0}}
\centerline{\psfig{figure=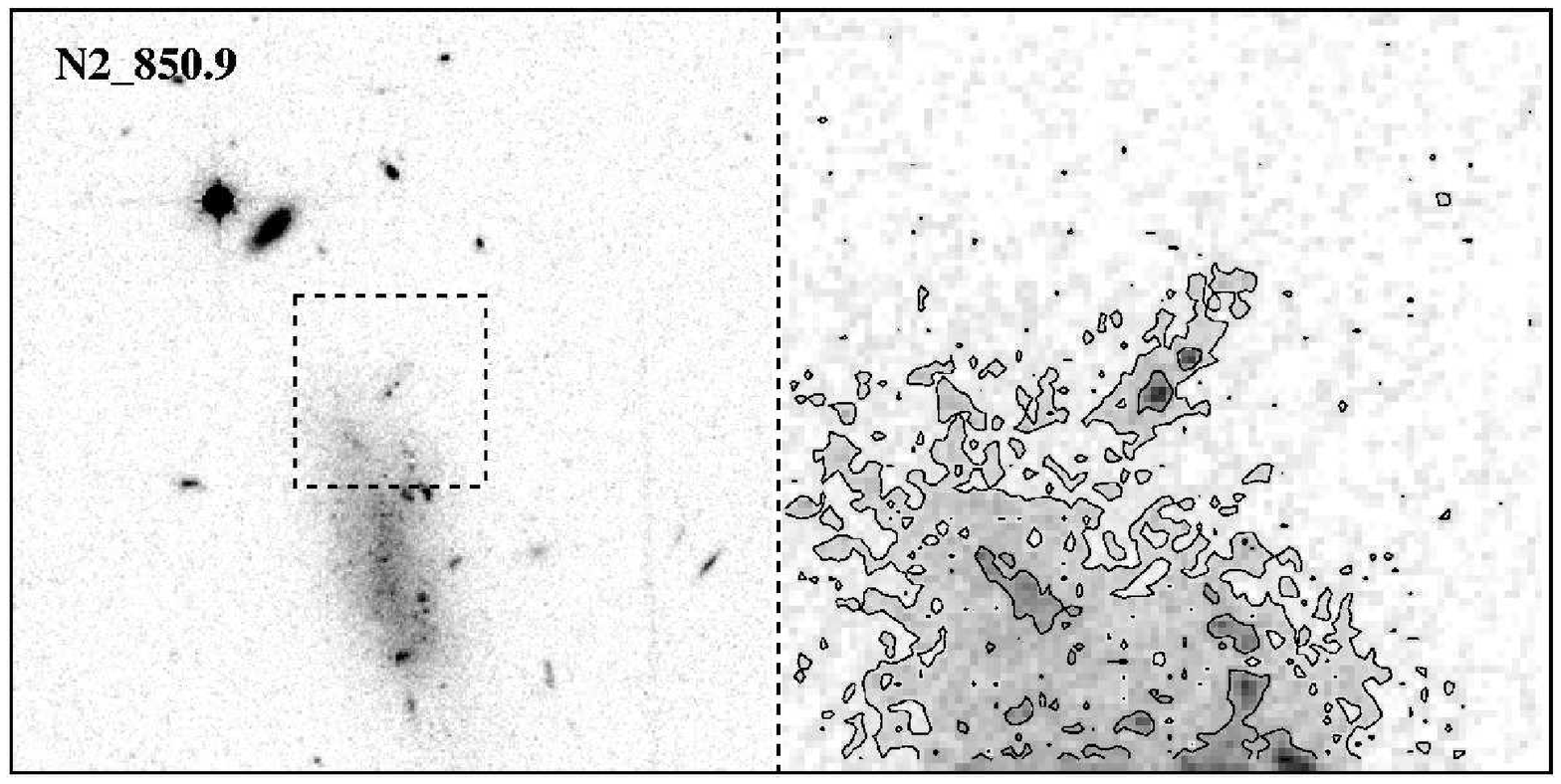,width=0.5\textwidth,angle=0}
\psfig{figure=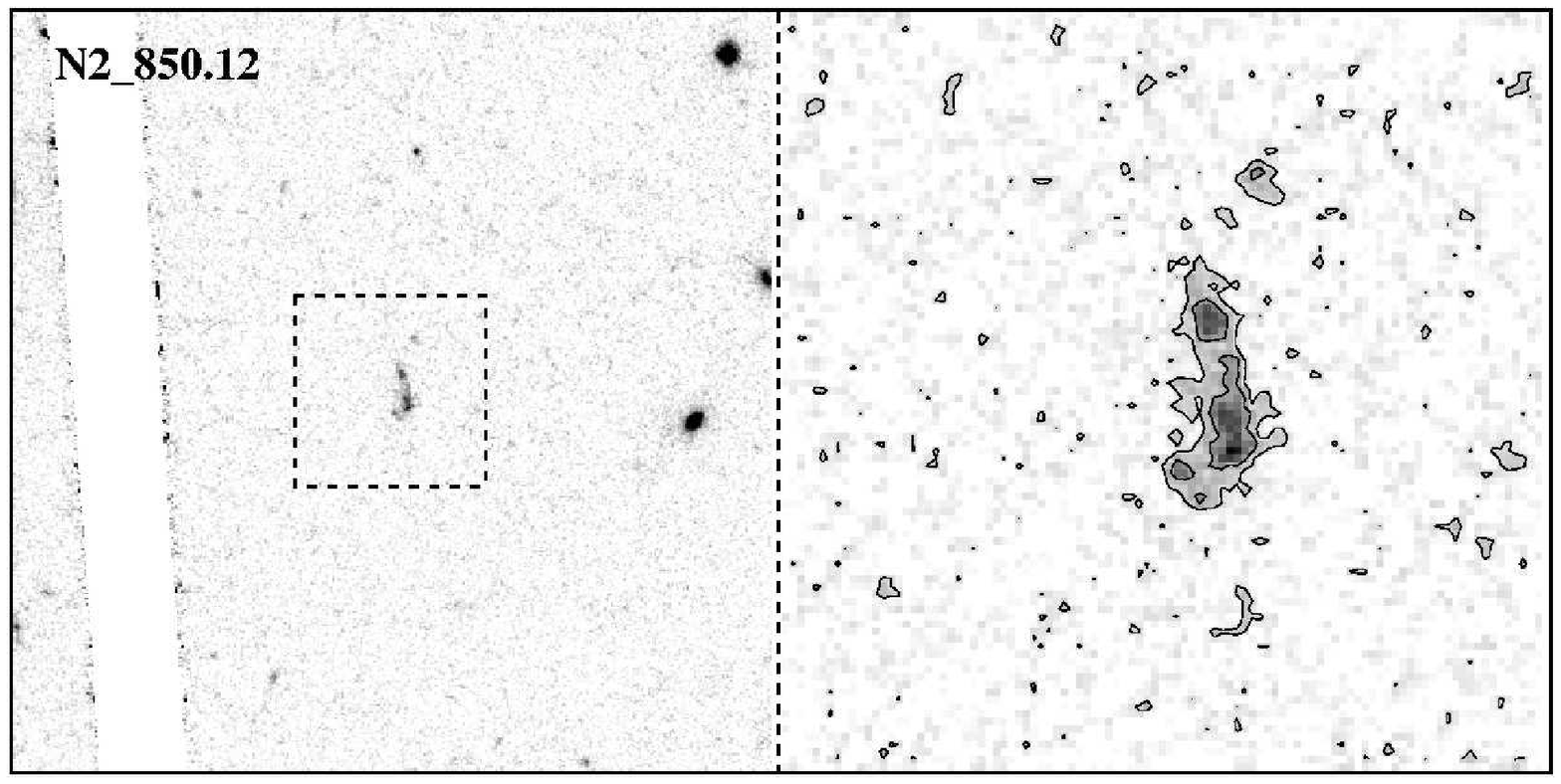,width=0.5\textwidth,angle=0}}
\caption{ HST-ACS images of the 4 submm sources with {\it likely}
optical counterparts (see Section 3), centred on galaxies which are
either ERO or VRO in $I-K$ colours.
The $20\times 20$
arcsec postage stamps (left panels) permit a comparison with Ivison et
al. (2002), with a $5\times 5$ arcsec zoom showing contours to reveal
low surface-brightness features. The contour are  separated by intervals
of 1 magnitude/arcsec$^2$. }
\end{figure*}

\section{The observational data}

\subsection{The sample of submm galaxies}

The sample of SCUBA galaxies is drawn from the study of Scott et
al. (2002), which covered 260 arcmin$^2$ over two fields (ELAIS-N2 and
the Lockman-Hole) using a jiggle-mapping technique to an rms
noise level of $\sim 2.5$mJy per beam at $850\mu$m. 

The catalogue of Scott et al. (2002) presents 17 sources in the
ELAIS-N2 region above a significance threshold of $3.5\sigma$.
Following Ivison et al. (2002), we exclude one source which lies in a
region where the noise is in excess of 3mJy per beam, and a further
source which was not confirmed in further SCUBA imaging. This leaves a
sample of 15 SCUBA submillimetre sources which form the basis of our
analysis.

\subsection{HST and Gemini imaging}

Observations with the HST Advanced Camera for Surveys (ACS) were taken
of the ELAIS-N2 field during Cycle 13\footnote{Obtained from the Space
Telescope Science Institute, Program ID \# 9671, PI Almaini}.  The
data form a mosaic of 8 fields to a depth of two orbits per pointing,
observed in the F814 filter. Exposure times were 4760s per tile, with
the exception of one region which was observed for 4284s to
accommodate a fixed roll angle (to avoid saturation from a bright
star). Exposures were taken in a standard 4-pt dither pattern split
over two orbits, using the DITHER-BOX configuration.  Data were
reduced, distortion corrected and re-drizzled onto a 0.05 arcsec pixel
frame using the standard IRAF STSDAS and MULTIDRIZZLE software.

Deep K-band images were obtained on Gemini North
\footnote{Program ID GN-2002A-Q-12, PI Dunlop} with the Near InfraRed
Imager (NIRI), using  an integration time of 90 minutes per
source. Observations were made in photometric conditions with seeing
$< 0.6$ arc-seconds, and consisted of a standard 9-point jitter
pattern with 10 arcsec offsets.  This reaches a point-source depth
of approximately $K=22$ (5$\sigma$).  Further details can be found in
Targett et al. (in preparation).

\section{Identification of optical candidates}

The identification of optical candidates for the submm
galaxies is presented in detail in Ivison et al. (2002). This is based
on deep radio imaging at 1.4GHz taken with the Very Large Array (VLA),
reaching a noise level of approximately 9$\mu$Jy per beam at a
resolution of 1.4 arcsec. This is combined with ground-based optical
and infrared imaging from the William Herschel Telescope (WHT) and the
UK Infrared Telescope (UKIRT), reaching $3\sigma$ point-source
limiting magnitudes of $V=25.9, R=26.0, I=25.0$, and infrared depths
in the range $K=20-21$.  Full details are provided in Ivison et
al. (2002), while a description of the near-infrared data is provided
in Roche et al. (2002). Chandra X-ray observations of this field are
discussed in detail in Manners et al. (2003) and Almaini et
al. (2003).

On the basis of these data, combined with our deeper $I-$ and $K-$band
imaging from HST and Gemini, we attempt to identify the most reliable
optical counterparts for further morphological study. We separate the
15 submm sources into 3 categories: those with `secure', `likely'
and `unknown' identifications.

Ivison et al. (2002) showed that 6 of the 15 sources have unambiguous
radio detections, in each case identified with a clear optical or
infrared counterpart. Given the low probability of a chance radio
alignment (typically $1-2$ per cent), these are designated `secure'
submm identifications. All  of these galaxies now have claimed spectroscopic
redshifts, as described in Section 6.

Of the remaining 9 submm sources, the lack of an unambiguous radio
detection makes the identification process far less certain. Ivison et
al. (2002) identify one or more plausible counterparts to 8 of these
sources, although the choice is often rather subjective.  For the
purposes of this paper we therefore follow Pope et al. (2004) and
impose the additional selection criteria that submm galaxies are
typically very red in $I-K$ (see also Smail et al. 2004).  We search
for such counterparts within 7 arcsec of the submm centroid,
corresponding to an approximate $2\sigma$ error circle (Ivison et
al. 2002).  Of the 9 remaining sources, we find that 2 contain
(single) galaxies that satisfy the definition of Extremely Red Objects
(EROs) with $(I-K)_{Vega}>4 $ and a further 2 contain `Very Red
Objects' (VROs), which we define as $(I-K)_{Vega}>3.5$.  We designate
these as `likely' identifications. Based on the space density of EROs
and VROs to the depth of our Gemini data ($K\simeq 21$), and the
submm/IR offsets of our candidates, we estimate an expectation value
of $\simeq 0.8$ spurious ERO/VRO associations.  This would suggest
that while most of the `likely' IDs are probably correct, there is a
good chance that some may be chance coincidences.  We will therefore
treat these separately from the 6 `secure' IDs in the analysis that
follows.

This process leaves 5 submm sources from our sample of 15 without any
radio or ERO/VRO counterpart.  Since we have no reliable indication of
the likely identifications in these cases (such as the coincidence
with a Chandra X-ray source) we remove these from the remainder of the
morphological analysis.  We discuss the potential biases this may
introduce in Section 7.

\begin{table*}  
\caption{Details of morphological parameters for sub-mm sources
illustrated in Figure 1 and Figure 2.  The concentration, asymmetry
and M\_{20} parameters have not been corrected for redshift. Redshifts
in bold are spectroscopic (Chapman et al. 2005), while the remainder
are photometric (Aretxaga et al., private communication).  Sizes in
parentheses are derived from K-band model fitting rather than the
I-band HST data (Targett et al. 2005).  } \normalsize
%\scriptsize
\label{tab:CAS}
\begin{tabular}{cccccccccc}
\hline
SCUBA ID	&  $I_{\rm{814}}$ mag	& $K_{\rm{mag}}$ & Redshift	& C  		& A 	   & M\_{\rm 20} & $r_h$(arcsec) & $r_h$(kpc)&  $r_{Petr}$ (arcsec)  \\
\hline	
N2\_850.1 &	22.0		& 19.5	& {\bf 0.845}	& 3.0~$\pm$~0.4	 & 0.16~$\pm$~0.03 	&	  -1.9 &  0.43	   & 3.35	&  1.26 \\ 	
N2\_850.2 &	$>$25.5		& 20.4	& {\bf 2.454}	& -		 & - 		   	&	  -  &  (0.20)	   & (1.66)	& (1.60) \\
N2\_850.3 &	24.8		& 21.1	&	4.0	&3.5~$\pm$~0.16  & 0.31~$\pm$~0.20 	&	-1.5 &  0.40 	   & 2.83	&  0.90  \\
N2\_850.4 &     21.8  		& 18.4	& {\bf 2.378}	& 2.6~$\pm$~0.40  & 0.63~$\pm$~0.02 	&	  -1.5 &  0.30	   & 2.80	&  1.07	\\
N2\_850.6 &     23.9  		& 19.8	&	3.0	& 2.6~$\pm$~0.40  & 0.24~$\pm$~0.11  	&	  -1.4 &  0.38	   & 2.85	& 1.14	\\
N2\_850.7 &     22.4  		& 19.5	& {\bf 1.488}	& 3.1~$\pm$~0.30  & 0.33~$\pm$~0.09  	&	  -1.6 &  0.54        & 4.55	& 1.75 	\\
N2\_850.8 &     21.6  		& 18.2	& {\bf 1.190}	& 3.6~$\pm$~0.30 & 0.50~$\pm$~0.03   	&	  -2.1 &  0.54	   & 4.45	& 1.82	\\
N2\_850.9 &     25.1  		& 20.8	& 3.4	& 2.5~$\pm$~0.30 & 0.16~$\pm$~0.11   		&	  -1.5 &  0.45	   & 3.36	& 1.82	\\
N2\_850.12 &    24.2	   	& 20.6	&    {\bf  2.425} & 3.1~$\pm$~0.40	& 0.16~$\pm$~0.11  &  	  -2.5 &  0.36	   & 2.96	& 1.53	\\
N2\_850.13 &    $>$25.5   	& 20.9	& {\bf 2.283}	& - 		 & - 		   	&	   -  &  (0.42)	   & (3.45)	& (1.32)	\\
\hline 
\normalsize
\end{tabular}
\end{table*}

\section{Quantitative morphological analysis}

Postage stamp HST images for our sample are shown in Figures 1 \& 2.
A visual inspection of the 6 secure submm galaxies (Figure 1) reveals
three sources with clear evidence of disturbed, possibly merging
morphology, one face-on disk system and two very faint ERO galaxies.
Of the 4 `likely' IDs (Figure 2) all appear to be faint compact
systems, with at least two showing evidence for disturbed morphologies
by eye. Most of the galaxies also appear very compact and sharply
peaked in their light distribution. 

To conduct a more quantitative assessment of their morphologies we
adopt the CAS (concentration, asymmetry and clumpiness) structural
analysis system developed by Conselice (2003). One advantage of this
technique is the ability to trace structural features that can be
related to past and present star formation and merger activity.  A
rest-frame optical/UV asymmetry value in excess of $A=0.35$ was found
to be a robust indication of a major merger at $z=0$ (Conselice et
al. 2003a), while the concentration index is found to be a fair
representation of scale of a galaxy and the fraction of stars in a
bulge component (Graham et al. 2001; Conselice 2003).

Two of the `secure' submm galaxies ($N2\_850.2$ and $N2\_850.13$) are
too faint to produce converging CAS parameters in our I-band HST
imaging. Both are ERO galaxies with $I_{\rm Vega}>25$. We therefore
exclude these galaxies from the morphological analysis, although we
present an examination of their K-band light profiles in Targett et
al. (in preparation).

\subsection{Structural comparison with local galaxy populations}

In Figure 3 the concentration/asymmetry parameters for our submm
galaxies are compared to those of local galaxy populations, following
the method of Conselice, Chapman \& Windhorst (2003b).

To perform a fair comparison with local galaxy populations it is first
important to determine how the measured morphology of a local galaxy
would change when shifted to the rest-frame UV and observed at
high-redshift with the ACS camera.  These corrections were determined
using simulations, artificially redshifting 50 nearby ULIRGs and 82
normal Hubble types to examine how they would appear in our ACS images
at redshifts $z=1-3$. The resolution and surface brightness are degraded, and
noise is added to match the real data. A full description of this
technique is outlined in Conselice (2003) and Conselice et
al. (2003b). Using these new images, CAS parameters are then
re-measured for the simulated galaxies, in the same manner as the
submm galaxies, in order to determine appropriate redshift corrections.

Based on these simulations, we can correct the CAS parameters of the
submm galaxies to those of local galaxies.  The most conservative
approach is to adopt the redshift corrections appropriate for normal
spiral galaxies, which change least in their CAS values after
redshifting (e.g. $\Delta A=0.09$, $\Delta C=-0.06$ at $z=2$ compared
to $z=0$).  For the three submm galaxies without spectroscopic
redshifts (all in the `likely' category) we use photometric redshifts
(see Section 6.2) but limit the C/A corrections to the maximum value
appropriate to $z=3$ ($\Delta A=0.21$, $\Delta C=-0.06$).  The
resulting comparison is shown in Figure 3.

The most striking finding is that the submm sources are significantly
more asymmetric than one would expect for local spiral or elliptical
galaxies, and are much more similar in concentration and asymmetry to
local ULIRGS.
A one-dimensional KS test along the asymmetry axis rejects the
null-hypothesis that submm sources are drawn from the populations of
local elliptical and spiral  galaxies at $>99.9$ and $>99.5$ per cent
significance respectively. The same comparison with ULIRG asymmetry values
yields no significant difference.

Using the major merger criteria calibrated in Conselice et al. (2003a)
for $z\sim 0$ galaxies ($A>0.35$) we find that $6/8$ are classified as
major mergers.  Allowing for the two galaxies which were too faint to
classify this becomes a minimum fraction of $6/10$.  We note that
these fractions are formed by taking the conservative assumption that
the CAS redshift corrections for the submm sources are the values
appropriate for normal galaxies.  Instead, if we assume the ULIRG
redshift correction then the submm asymmetry values shown in Figure 3
would increase by $\Delta A \simeq 0.2$, and the merger fraction would
be substantially higher. We conclude that {\em at least} 60 per cent
of the submm galaxies with secure/likely IDs are classified as
mergers.

\begin{figure}
\psfig{figure=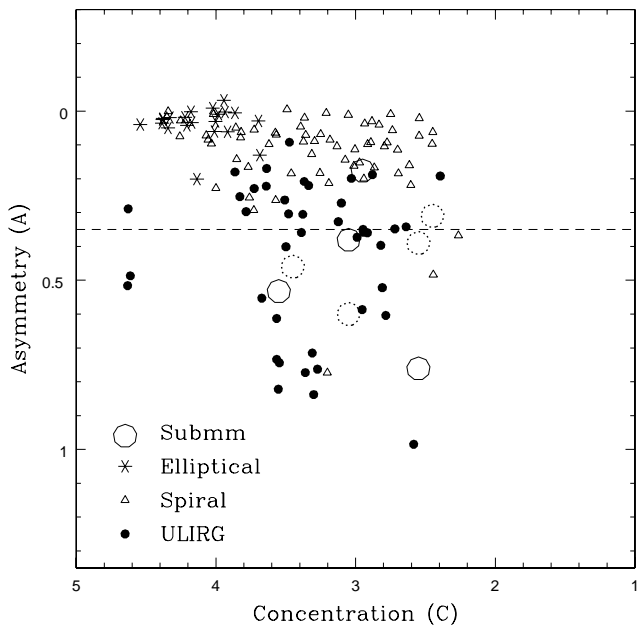,width=0.5\textwidth,angle=0}
\caption{Concentration-asymmetry diagram for samples of spiral
galaxies, elliptical galaxies and ULIRGs (from Conselice et al. 2000),
compared with values for the submm sources studied in this paper.  The
submm C/A values have been corrected to allow for the effects of
redshifting and observation in the F814 filter, following simulations
described in Section 4. The broken submm source symbols are those with
only `likely' optical identifications. Galaxies are classified as
major mergers if $A>0.35$ (Conselice et al. 2003). }
\end{figure}

\subsection{Comparison with Lyman-break galaxies}

\begin{figure}
\psfig{figure=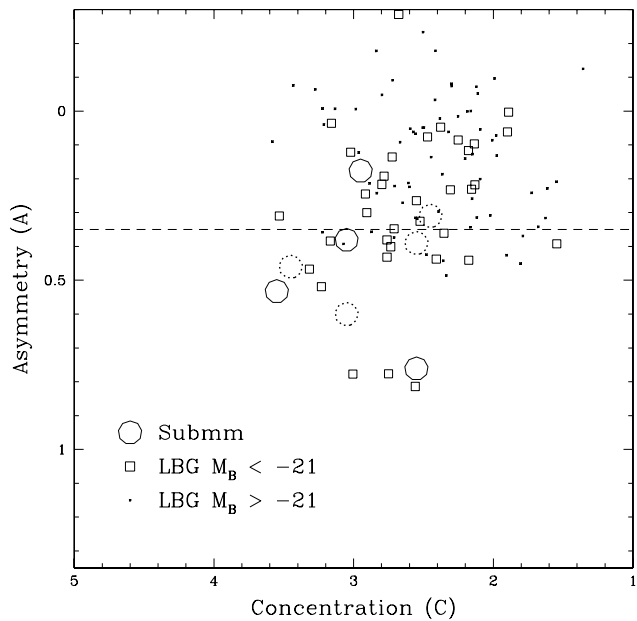,width=0.5\textwidth,angle=0}
\caption{Concentration-asymmetry diagram for our submm galaxies
compared with Lyman-break galaxies in the redshift range
$2<z<4$ (observed using WFPC2 in the HDF).  All C/A values have been
corrected to allow for the effects of redshifting and observation in
the respective F814 filters. The broken submm source symbols are those
with only `likely'  optical identifications.  The Lyman-break galaxies
are separated into two luminosity bins. }
\end{figure}

While it is clear that the submm galaxies are morphologically distinct
from normal low-redshift galaxies, it is also valuable to perform a
comparison with `normal' galaxies at more comparable redshifts.  Such
galaxies are arguably the Lyman-break galaxy (LBG) population, first
detected in large numbers in the pioneering work of Steidel et
al. (1996).

In Figure 4 we compare the concentration/asymmetry parameters for the
submm galaxies with those of photometrically-selected galaxies in the
Hubble Deep Field North (HDF-N) in the redshift range $2<z<4$, for
which CAS parameters have already been determined in the work of
Conselice et al. (2003). We find that submm galaxies show a tendency
towards higher concentrations and greater asymmetries. A
Kolmogorov-Smirnov (KS) test can reject the null hypothesis that they
are drawn from the same distribution in asymmetry with a confidence
level of $>95$ per cent.  In concentration, the KS test can reject the
corresponding null hypothesis more strongly, at $>99$ per cent
confidence.

Since the LBGs cover a wide range in luminosity ($-22.6 < M_B <
-18.5$), we separate these into high and low luminosity sub-samples
(at $M_B=-21$) and repeat the KS comparison with the submm galaxies.
We find that the difference in asymmetry is largely due to the
low-luminosity LBGs, and formally do not find a significant difference
with the high-luminosity sub-sample. This is consistent with the
findings of Conselice, Chapman \& Windhorst (2003), where it was found
that the major merger fraction among submm sources is similar to that
of the most luminous Lyman-break galaxies. Interestingly, however, the
difference in {\it concentration} distributions remains similar for
both LBG sub-samples (significant at $>95$ per cent in each case).

\subsection{$M\_{20}$ analysis}

A new morphology estimator, the $M_{\rm 20}$ index, has recently been
developed by Lotz, Primack \& Madau (2004). This is defined as the
second-order moment of the brightest 20 per cent of light, and was
found to be a sensitive indicator of merger signatures such as
multiple nuclei.  For completeness, we include measurements of the
$M_{\rm 20}$ index in Table 1. We find a broad range of values for the
submm galaxies, with a median value of $M_{\rm 20} = -1.55$.  By
comparison with the study of Lotz, Primack \& Madau (2004), we find
that these are intermediate between those of normal galaxies
($\overline{M}_{\rm 20}\simeq -2.0$) and ULIRGs ($\overline{M}_{\rm
20}\simeq -1.5$). This suggests a mixture of morphological types, with
a tendency for submm galaxies to have morphologies more like ULIRGs
than normal local galaxies, which is consistent with the findings of
the C/A analysis given above.

\section{Half-light radii and physical sizes}

\subsection{Measuring galaxies sizes}

The process of measuring size scales for faint galaxies can be prone
to a number of systematic biases.  Traditional methods use the radius
at which a certain isophotal surface-brightness is reached, but these
can be prone to uncertainties in the photometric zero-point.  Such
measurements are also not robust for studying faint and/or low-surface
brightness objects, particularly when comparisons are made over a wide
range in redshift, due to the strong $(1+z)^4$ surface-brightness
dimming.  We have therefore adopted the technique of Bershady et
al. (2000), which uses a non-isophotal method to define the
total-aperture magnitude, independent of the surface-brightness
distribution and the photometric calibration.  This is based on a
dimensionless parameter $\eta$ which is defined as the ratio of the
average surface brightness within radius $r$ to the local surface
brightness: $\eta(r) \equiv I(r)/\langle I(r) \rangle$.  We define the
Petrosian radius $r_{Petr} = 1.5 \times r_{(\eta=0.2)}$, following
Conselice (2003). The half-light radius ($r_h$) is then obtained at the
radius which encloses half of this total-light value.  

For comparison we also measure  half-light radii using a standard
curve-of-growth technique. This was found to give virtually identical
values to the method outlined above (typically within $\pm 1$ pixel).
As a third test, we also obtained half-light radii using the output
from the SExtractor software (Bertin \& Arnouts 1996). In some cases
this required some tuning of the deblending parameters to prevent the
software from separating multi-component features into separate
objects. These resulting values for $r_h$ were found to be in
generally good agreement with the other methods, although on average
$0.05$ arcsec {\em smaller}. We believe this  difference is due
to the SExtractor software measuring features within a segmentation
map which effectively truncates the outer regions of a galaxy.  We
stress, however,  that this  small difference does not affect any of the
conclusions presented here.

Since these techniques are broadly consistent, hereafter we adopt the
half-light and Petrosian radii measured using the first method.  These
are  presented in Table 1.

\begin{figure}
\psfig{figure=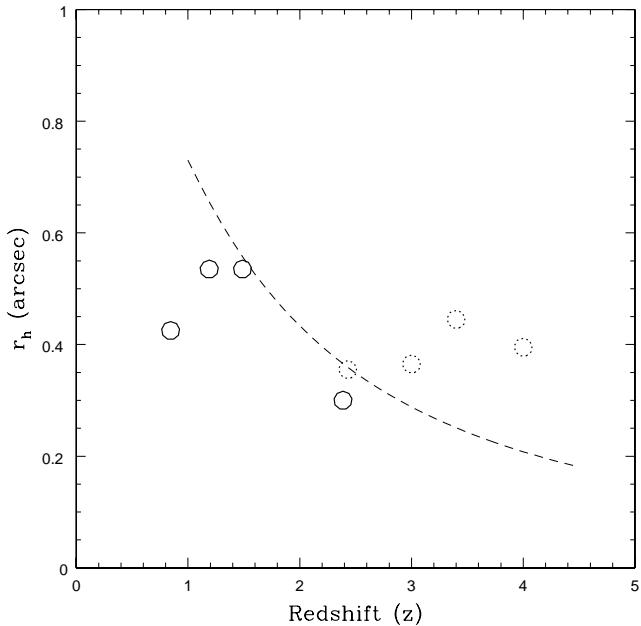,width=0.5\textwidth,angle=0}
\caption{Size vs. redshift relation for the SCUBA galaxies, with
broken symbols reflecting those with only `likely' optical
identifications.   For
comparison, the curve shown is taken from Ferguson et al. (2004),
which was found to provide a good description of normal high-redshift
field galaxies from the GOODS surveys, evolving as $H^{-1}(z)$.}
\end{figure}

\subsection{Comparison with field galaxies and  LBGs}

\begin{figure}
\psfig{figure=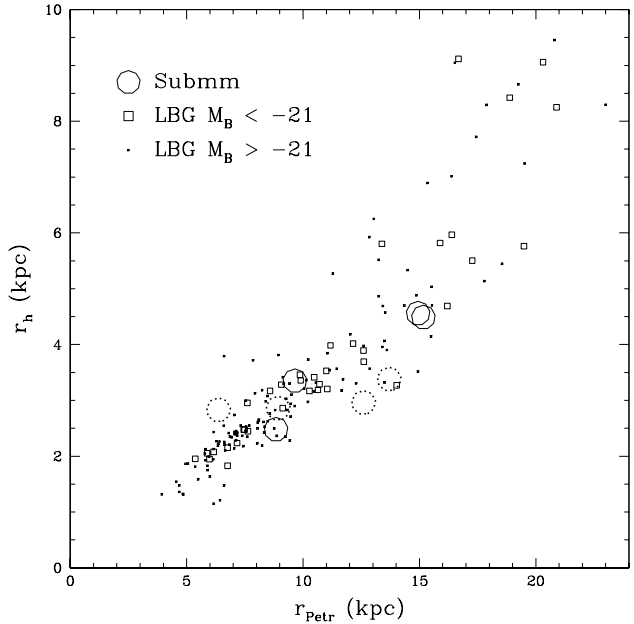,width=0.5\textwidth,angle=0}
\caption{Comparison of submm and Lyman-break galaxy sizes, showing
both half-light and Petrosian radii. The broken submm source symbols
are those with only `likely' optical identifications.  The Lyman-break
galaxies are separated into two luminosity bins.  }
\end{figure}

In Figure 5 we plot submm galaxy sizes as a function of redshift.  For
the three galaxies without spectroscopic redshifts (all in the
`likely' category) we use photometric redshifts provided by
I. Aretxaga (private communication; see Section 6.2). For comparison,
the curve shows the mean half-light radii for field galaxies as a
function of redshift from the study of Ferguson et al. (2004), which
is based on HST ACS observations of the GOODS deep fields.  Galaxies
in this study were restricted to rest-frame luminosities in the range
$0.7-5L^*$ (where $L^*$ is the characteristic luminosity of a $z=3$
LBG from Steidel et al. 1999). They found that the mean size evolution
can be well described by a function of the form $H^{-1}(z)$, where
$H(z)$ is the Hubble parameter.  Although our submm sample is small,
it is notable that we see no clear difference in average size compared
to the LBG population (Figure 5).  There is marginal evidence that the
highest redshift ($z>3$) submm galaxies lie above the LBG relation,
but since these all have uncertain optical IDs it is impossible to
draw firm conclusions.

As a second test, in Figure 6 we compare both half-light and Petrosian
radii with the LBG study of Conselice et al. (2003b), the same sample
used in the morphological comparison in Figure 4.  This has the
advantage that we are evaluating the $r_{Petr}$ and $r_h$ values in
{\em precisely} the same manner as the submm galaxies, and all based
on observations in the HST F814 filter. Since the LBGs cover a wide
range in luminosity ($-22.6 < M_B < -18.5$) we once again split the
sample into two subsets.  In terms of both half-light and Petrosian
radii, the submm galaxies are consistent with the distribution of
sizes seen in either  class of LBG (as confirmed by a a KS test).
For the most luminous
LBGs with (absolute magnitudes $M_B<-21$) we find a mean half-light
radius of $r_h=4.6\pm3.2$~kpc, while the fainter sample give
$r_h=3.2\pm1.7$~kpc (where errors reflect the standard deviation, not
the error on the mean).  For comparison, our submm galaxies yield
$r_h=3.2\pm0.84$~kpc. Restricting the sample to the `secure' IDs, we
obtain $r_h=3.8\pm0.85$~kpc. We conclude that the submm galaxies have
mean radii that are between those of faint and luminous LBGs,
although statistically they are indistinguishable from either class.

\subsection{Comparison with previous studies of submm galaxies}

At face value, our results appear to be in contrast to previous
studies (Chapman et al. 2003b, Smail et al. 2004, Pope et al. 2004)
which report that submm galaxies are typically larger than LBGs,
perhaps by up to a factor of two. In this section we briefly compare
our findings with these previous papers.

Our distribution certainly differs from the study of Smail et
al. (2004), who present sizes for $15$ galaxies in the GOODS-N field.
(a mixture of submm galaxies and optically-faint radio galaxies). Of
these, they report $7$ with half-light radii in excess of $0.8$
arcsec. In contrast, none of the galaxies in our study have $r_h$
values exceeding $0.6$ arcsec. According to a KS test, the probability
that our galaxies are drawn from the same underlying distribution is
less than 1 per cent, suggesting a clear difference between our samples.

Chapman et al. (2003b) present a study of $12$ submm galaxies, and
both half-light and Petrosian radii are tabulated.  According to KS
tests, our samples are statistically indistinguishable using either
radius measurement. The only notable difference is the mean value for
the Petrosian radii, which is $2.1$ arcsec in their work compared to
$1.4$ arcsec from our sample. In Chapman et al. (2003b) their large
mean Petrosian value is compared to the mean value for LBGs in the
HDF-N ($1.2$ arcsec) as evidence that submm galaxies are larger. We
note, however, that this mean value is dominated by two galaxies which
are substantially larger than the rest in their sample, with Petrosian
radii of $7.63$ and $4.05$ arcsec.  Both are multiple-component
systems, one of which is also in the sample of Smail et al.  2004. If
instead we compare {\em median} Petrosian radii, the Chapman et al.
sample yields $1.22$ arcsec, which is in good agreement with the
median value for LBGs and formally smaller than the median value for
our sample ($1.40$ arcsec).

Pope et al. (2004) quote only Petrosian radii, and again a KS test
reveals no statistically significant difference compared to our
study. The mean value from their sample is $1.12$ arcsec, which is
marginally smaller than our sample mean of $1.41$ arcsec.  The GOODS
HST data used in their paper is deeper than our observations, so a
fairer comparison would be to restrict the Pope et al. sample to those
with $i(AB)_{775}<26$, which is comparable to the limits of our data.
A KS test again reveals no difference, although formally their mean
value is now closer to ours at $1.33$ arcsec.  No comparison with LBGs
is made, but it is reported that submm galaxies are generally larger
than field galaxies in the HDF-N. However, this conclusion is based on
a sub-sample (those which are either optically-bright and/or at
$z<2$), and has a statistical significance of only $2\sigma$.

In summary, we conclude that the submm galaxy radii measured in
previous studies are broadly consistent with our findings.  Only the
work of Smail et al. (2004) presents galaxies which are significantly
larger than our findings. We also note that the previous studies have
strongly overlapping samples.  We therefore suggest that the perceived
wisdom that submm galaxies are larger than LBGs is so far
inconclusive. A fairer statement may be that submm galaxies show a
wider {\em variance} in properties compared to LBGs, but further
(independent) studies are clearly required before firm conclusions can
be drawn.

\section{Notes on individual sources}

These brief notes are designed to complement the descriptions given in
Ivison et al. (2002), where further information on these submm
galaxies can be obtained.

\subsection{Secure identifications}

{\bf N2\_850.1}: This is the brightest submm source in the field,
associated with a close pair of radio sources. A relatively bright,
compact galaxy is aligned with the brightest knot of radio emission,
which was subsequently found to lie at a redshift $z=0.840$ (Chapman
et al. 2005).  The 450/850$\mu$m flux ratios would then imply a
surprisingly low dust temperature ($T_{dust}\sim 23$K at $z=0.84$), so
it has been suggested that this could be evidence for strong
gravitational lensing by a massive, compact foreground galaxy (see
Ivison et al. 2002, Chapman et al. 2002). Our observations would
appear to rule out this strong-lensing hypothesis, since the HST
imaging reveals this to be a clear face-on spiral galaxy (which was not
clear from ground-based data). The integrated mass density from a
face-on spiral would not be sufficient to cause strong gravitational
lensing (Moller \& Blain 1998). However, we note that a faint knot of
optical emission is resolved at approximately the location of the
weaker radio source, approximately 1 arcsec east of the galaxy centre.
While more likely to be a knot of star-formation in a spiral arm, this
could plausibly be associated with a more distant background
object. There is therefore a slim possibility of a chance alignment
with the bright foreground galaxy, although we estimate the
probability of a chance coincidence with a radio-emitting galaxy of
this optical magnitude to be $<<1$ per
cent. We note, however, that this is a region of strongly enhanced
foreground galaxy density, where the probability of a chance alignment
could be enhanced by the more subtle effects of weak gravitational
lensing bias (Almaini et al. 2005).

{\bf N2\_850.2}: This submm source is associated with a close pair of
ERO galaxies. The northern (optically-fainter) ERO is
associated with the radio emission and hence the likely source of
submm emission.  Chapman et al. (2005) find a redshift $z=2.454$

{\bf N2\_850.4}: This is a well-studied submm galaxy at $z=2.378$,
examined in detail by Smail et al. (2003) and Swinbank et al. (2004).
HST reveals a clearly disturbed multiple-component morphology
(formally the most asymmetric galaxy in the sample; see Table 1).
Spatially resolved spectroscopy shows evidence for vigorous starburst
activity, an active Seyfert component and a galactic-scale starburst
driven wind (Swinbank et al. 2004).

{\bf N2\_850.7}: HST reveals an unambiguous merger system, with a
highly distorted morphology and possibly  3 distinct
components. Chapman et al. (2004) find a redshift $z=1.488$.

{\bf N2\_850.8}: This is the only submm source detected in X-rays by
Chandra (Almaini et al. 2003). The X-ray luminosity and spectrum
suggest an absorbed, moderate luminosity AGN.  Ground-based imaging
indicated only a very compact galaxy, but this is spectacularly
revealed by HST as a face-on spiral with a bright, highly distorted
nucleus.  At least two compact nuclei can be distinguished, strongly
suggesting a recent major merger.  Chapman et al. (2005) find a redshift of
$z=1.190$. Some association with the nearby ring galaxy seems likely,
which is confirmed by Chapman et al. (2005) to lie at the same
redshift.

{\bf N2\_850.13}: A strong radio source is associated with an
optically faint ERO, barely detected by HST but with a distorted
K-band morphology (Targett et al., in preparation). Chapman et
al. (2005) report a redshift $z=2.283$ and classify this as an AGN.

\subsection{Likely  identifications}

For completeness, we include brief notes on the galaxies classified as
`likely' identifications (see Section 3), although we stress that
these should be treated with some caution. Three of these galaxies do
not have a spectroscopic redshift, so we use photometric redshifts
from Aretxaga et al. (private communication; see also Aretxaga et
al. 2005).  These are based on a combination of radio, mm and submm
photometry, and do not depend on the uncertain optical/IR
counterparts.

{\bf N2\_850.3} A faint red galaxy, classified as a VRO ($I-K=3.7$).
A photometric redshift from Aretxaga et al. (private communication)
suggests $z\simeq 4.0$.  Our HST ACS imaging reveals a compact,
asymmetric light profile.

{\bf N2\_850.6} A very red galaxy, classified as an ERO ($I-K=4.1$),
with a photometric redshift estimate of $z\simeq 3.0$ (Aretxaga et
al., private communication).  HST reveals a distorted,
multiple-component morphology.

{\bf N2\_850.9} This source is in the `likely' category since there
are two possible radio counterparts to the submm emission.  One is
associated with a low-redshift, low surface-brightness (LSB) galaxy,
and the other is associated with a faint, red ERO to the north, with
extreme colours ($I-K=4.3$). Ivison et al. (2002) choose the LSB
galaxy as a `plausible' ID, while in this paper we favour the faint
ERO. We select the ERO following the argument outlined in Section 3,
noting also that this object clearly stands out in the K-band image
(where the LSB is barely detected). Furthermore, the radio/mm/submm
photometric redshift yields $z\simeq 3.4$ (Aretxaga et al., in
private communication) which would not be consistent with the LSB galaxy.

{\bf N2\_850.12} An elongated irregular galaxy, identified as VRO
($I-K=3.8$). This is also a weak radio source, although not a formally
significant radio-submm association (Ivison et
al. 2002). Near-infrared spectroscopy reveals a redshift $z=2.425$
(Simpson et al. 2004), which is consistent with a photometric
estimate of $z=2.5$ from Aretxaga et al. (2005).

\section{Summary and Conclusions}

We present high-resolution HST ACS imaging of a sample of 10 SCUBA
submillimetre galaxies. These are drawn from an initial flux-limited
sample of 15 submm sources from a contiguous field in the ELAIS-N2
region (Scott et al. 2002). We find that we can assign secure
optical/IR identifications to 6 submm sources with unambiguous radio
detections, and `likely' counterparts to a further 4 which are
associated with ERO/VRO galaxies.

A large fraction of the submm galaxies show disturbed, compact,
multi-component morphologies. In many cases these distorted
morphologies were not revealed with ground-based imaging. Using the
CAS system of quantitative morphological classification, we find that
at least $6$ are classified as major mergers.
Simulations suggest that the morphological parameters are very
different in appearance to local spiral and elliptical galaxies, in
particular by being significantly more asymmetric in their light
distribution. The morphological parameters are, however,  similar to those
of local ULIRGs.

A comparison with typical high-z field galaxies (the Lyman-break
population) shows that the submm galaxies are on average more
asymmetric and also significantly more concentrated in their light
distribution.  This difference in asymmetry drops if we restrict the
comparison to the most luminous Lyman-break galaxies, but the
difference in concentration remains and appears to be a clear
difference between these populations.

The submm galaxies display a narrow range of half-light radii in the
range $r_h = 2.8-4.6$~kpc. These sizes are consistent with
measurements of `typical' LBGs at $z=2-4$, and we reach the same
conclusion using Petrosian radii.  We therefore cannot support
previous claims that submm galaxies are significantly larger than LBGs
(Chapman et al. 2003b, Smail et al. 2004, Pope et al. 2005). We find
that the sizes reported  in previous studies are broadly consistent
with ours, so we suggest that the perceived wisdom that submm galaxies
are larger than LBGs is far from conclusive.

We note that we have been forced to exclude 5 submm sources without
reliable optical/IR counterparts, either because they were not
detected as radio sources or as ERO/VRO galaxies to the limit of our
optical/IR data. A further 2 were excluded from the morphological
analysis since they were too faint to give meaningful values.  These
effects are likely to bias us against the highest redshift systems, so
formally our conclusions are restricted to the optically-bright
($I<26$) low-redshift ($z \simlt 3$) submm galaxies. 

In summary, our observations lend support to the growing realisation
that many submm sources are triggered by major mergers.  These
galaxies also appear morphologically distinct from more typical
Lyman-break galaxies, with notably more concentrated light profiles.
We speculate that this could be due to a higher fraction of stellar
mass in the form of a central spheroidal component.

\section*{ACKNOWLEDGMENTS}

We are grateful to Alfonso Arag\'on-Salamanca, Meghan Gray, Kyle Lane
and Jennifer Lotz for useful discussions. We are also indebted to
Itziar Aretxaga for providing photometric redshifts in advance of
publication.  OA and RJM acknowledge the generous support of the Royal
Society.  TAT acknowledges the award of a PPARC studentship.  This
work is based on observations made with the NASA/ESA Hubble Space
Telescope, obtained from the data archive at the Space Telescope
Science Institute. STScI is operated by the association of
Universities for Research in Astronomy, Inc. under the NASA contract
NAS 5-26555.  This work is also based on observations obtained at the
Gemini Observatory, which is operated by the Association of
Universities for Research in Astronomy, Inc., under a cooperative
agreement with the NSF on behalf of the Gemini partnership: the
National Science Foundation (United States), the Particle Physics and
Astronomy Research Council (United Kingdom), the National Research
Council (Canada), CONICYT (Chile), the Australian Research Council
(Australia), CNPq (Brazil) and CONICET (Argentina).

\end{document}